# Adaptive threshold-based decision for efficient hybrid deflection and retransmission scheme in OBS networks

Martin Lévesque, Halima Elbiaze and Wael Hosny Fouad Aly
Department of Computer Science
Université du Québec à Montréal
Montréal (QC), Canada
Email: levesque.martin.6@courrier.uqam.ca, {elbiaze.halima, aly_w}@uqam.ca

*Abstract*—Burst contention is a well-known challenging problem in *Optical Burst Switching* (OBS) networks. Deflection routing is used to resolve contention. Burst retransmission is used to reduce the *Burst Loss Ratio* (BLR) by retransmitting dropped bursts. Previous works show that combining deflection and retransmission outperforms both pure deflection and pure retransmission approaches. This paper proposes a new *Adaptive Hybrid Deflection and Retransmission* (AHDR) approach that dynamically combines deflection and retransmission approaches based on network conditions such as BLR and link utilization. *Network Simulator 2* (ns-2) is used to simulate the proposed approach on different network topologies. Simulation results show that the proposed approach outperforms static approaches in terms of BLR by using an adaptive decision threshold.

## I. INTRODUCTION

*Optical Burst Switching* (OBS) [1] is a promising technology to handling bursty and dynamic Internet Protocol traffic in optical networks effectively.

In OBS networks, user data (IP for example) is assembled as a huge segment called a *data burst* which is sent using *one-way resource reservation*. The burst is preceded in time by a control packet, called *Burst Header Packet* (BHP), which is sent on a separate control wavelength and requests resource allocation at each switch. When the control packet arrives at a switch, the capacity is reserved in the cross-connect for the burst. If the needed capacity can be reserved at a given time, the burst can then pass through the cross-connect without the need of buffering or processing in the electronic domain.

Since data bursts and control packets are sent out without waiting for an acknowledgment, the burst could be dropped due to resource contention or to insufficient offset time if the burst catches up the control packet. Thus, it is clear that burst contention resolution approaches play an essential role to reduce the *Burst Loss Ratio* (BLR) in OBS networks [2].

Burst contention can be resolved using several approaches, such as *wavelength conversion*, *buffering* based on *fiber delay line* (FDL) or *deflection routing*. Another approach, called *burst segmentation*, resolves contention by dividing the contended burst into smaller parts called *segments*, so that a segment is dropped rather than the entire burst.

Deflection routing is an attractive solution to resolve the contention in OBS networks because it does not need added cost in terms of physical components and uses the available spectral domain. However, as the load increases, deflection routing could lead to performance degradation and network instability [3]. Since deflection cannot eradicate the burst loss, retransmission at the OBS layer has been suggested by Torra et al. [4].

A static combination of deflection and retransmission has been proposed by Son-Hong Ngo et. al. [5]. They have proposed a *Hybrid Deflection and Retransmission* (HDR) algorithm [5] which combines deflection routing and retransmission. Simulation results show that HDR gives bad performance in terms of BLR because it systematically try deflection first even if the load is high. To overcome this shortcoming, the authors have developed another mechanism called *Limited Hybrid Deflection and Retransmission* (LHDR) that limits the deflection.

This paper introduces a novel algorithm to combine deflection routing and retransmission called *Adaptive Hybrid Deflection and Retransmission* (AHDR). An adaptive decision threshold is used to dynamically make the decision of using either a deflection or a retransmission based on local knowledge about network conditions. The offset time is also adapted by using the adaptive decision threshold. In order to make the local knowledge feasible, AHDR algorithm exploits sending and receiving of *Positive Acknowledgement* (ACK) and *Negative Acknowledgement* (NACK) messages to advertise useful statistics about network conditions stored by all nodes.

This paper is organized as follows. Section II describes the proposed Adaptive Hybrid Deflection and Retransmission (AHDR) algorithm. Section III presents simulation results. Finally, Section IV contains the conclusion and future work.

## II. ADAPTIVE HYBRID DEFLECTION AND RETRANSMISSION

In this section, we describe the proposed algorithm (AHDR). AHDR optimizes the decision of performing either a deflection or a retransmission. When no contention occurs, the primary path is used (Fig. 1). However, when contention occurs, AHDR selects the best contention resolution strategy among deflection routing and retransmission (Fig. 2).

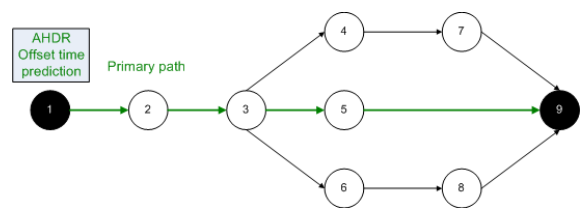

Fig. 1. Normal scenario, no congestion

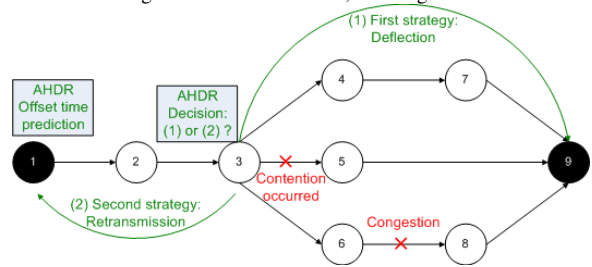

Fig. 2. Congestion scenario, contention occured

AHDR also enhances the selection of alternate routes. The offset time is determined at the ingress node where it is predicted by considering if a deflection will be needed or not.

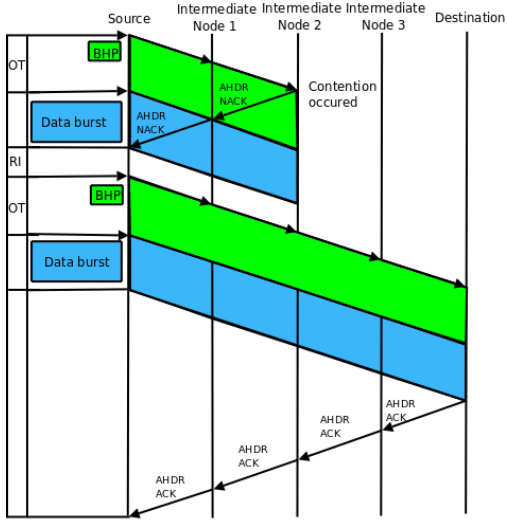

Fig. 3. Signaling scheme used by AHDR with a retransmitted burst

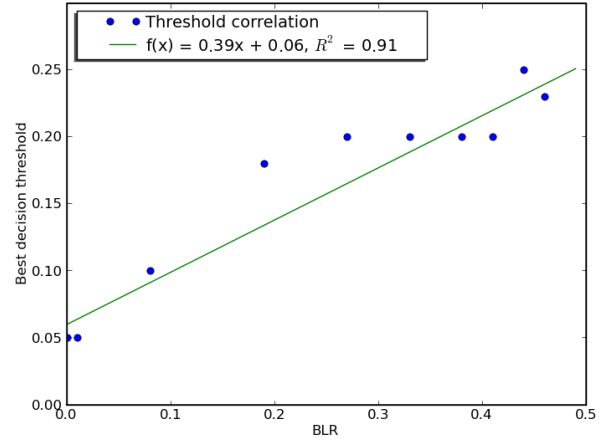

Fig. 4. Linear correlation between the BLR and the decision threshold

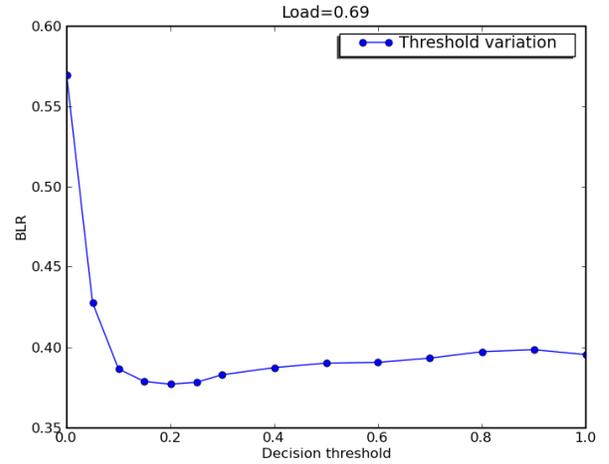

Fig. 5. Decision threshold variation with a fixed load of 0.69 in NSFNET

### A. Transferring statistics over the network

Once the control packet reaches the destination, an ACK is sent to the source. If the control packet is dropped, then the proposed algorithm uses a NACK to notify the source for burst retransmission. AHDR does not only use the ACK and the NACK for notifications but it uses them also to transmit some statistics about links states (Fig. 3).

The BLR and the link utilization are measured on each link and this information is integrated into notification packets. In the case of a NACK, statistics between the current node and the next node are used. In the case of an ACK, the BLR and the link utilization between the destination node and the last node before the destination are used. When a node receives an ACK or a NACK control packet, it collects and analyzes statistics. Thus, statistics of the whole network are eventually updated.

### B. Adaptive decision threshold

In order to resolve contention, the main idea is to decide whether to deflect or retransmit the burst. As the load increases, deflection routing can destabilize the network [3]. However, we want to maximize the bandwidth utilization by using as much deflections as possible without destabilizing the network by comparing a metric value and a certain decision threshold:

$$if\ (Metric\ Value \geq Decision\ threshold)\ then\ Deflect \quad (1)$$

In this paper the $Metric\ Value$ used is a success probability calculated with the BLR and the link utilization. Plus, $Decision\ threshold$ has to be adapted depending on network conditions as we will show later.

First, to limit the length of deflection routes, we introduce a parameter (noted $\xi$) which expresses the deflection route length threshold. Let $Defl$ denotes a possible deflection route, $Primary$ the primary route and $|Route|$ the number of hops of the route $Route$. If $|Defl| <= |Primary| * \xi$, then $Defl$ is added as a possible deflection alternative.

Second, AHDR incorporates BLR and link utilization weights to measure the *dropping probability* (DP) between two nodes as follows:

$$DP(n_1, n_2) = \alpha_{BLR} * BLR_{(n_1,n_2)} + \alpha_U * U_{(n_1,n_2)} \quad (2)$$

where $DP$ returns the dropping probability between $n_1$ and $n_2$, $U$ is the link utilization, $n_1$ and $n_2$ are two adjacent nodes, $\alpha_{BLR}$ is a weight applied to the BLR and $\alpha_U$ is another weight applied to the link utilization. We note that $\alpha_{BLR} + \alpha_U = 1$.

The success probability ($SP(R)$) of a route $R$ is defined as follows:

$$SP(R) = \prod_{i=1}^{|R|-1} (1 - DP(n_i, n_{i+1})) \quad (3)$$

The success probability of the link between $n_i$ and $n_{i+1}$ is given by $1 - DP(n_i, n_{i+1})$. Eq. 3 multiplies all success probability links to get a global success probability for the entire route.

In AHDR, an adaptive decision threshold is used to make the decision of either deflecting or retransmitting a given burst in contention.

Fig. 4 shows the linear correlation in *NSF Network* (NSFNET) between the BLR and the decision threshold. As the BLR increases, the decision threshold must also increase in order to reduce the number of deflections. All permutations of several loads and various decision thresholds were tried during the experiments. One point in the graph corresponds to the couple (BLR, Decision threshold) of the best decision threshold of a given load. For example, if we fix the load to 0.69 (Fig. 5), we note that the best decision threshold is near 0.2. If we use the regression line (Fig. 4) with $BLR = 0.37$, we get $f(0.37) = 0.2043$, which is near 0.2. The squared correlation coefficient ($R^2$) [6] quantifies the goodness of fit and is defined as follows:

$$R^2 = \left( \frac{\frac{1}{N-1} \sum_{i=0}^{N} BLR_i * th_i - N * \overline{BLR} * \overline{th}}{s_{BLR} * s_{th}} \right)^2 = 91\% \quad (4)$$

$R^2 = 91\%$ indicates that the correlation between the BLR and the best decision threshold for an effective decision is very high.

In order to take decisions by considering network conditions, we introduce a decision threshold function ($SP_{th}$):

$$SP_{th}(BLR_{Topo}, U_{Topo}) = \beta_{BLR} * BLR_{Topo} + \beta_U * U_{Topo} \quad (5)$$

where $\beta_{BLR}$ is a weight applied to the BLR of the entire network (known locally) and $\beta_U$ is a weight applied to the link utilization of the network. We note that $\beta_{BLR} + \beta_U \leq 1$.

If the success probability (Eq. 3) of a given deflection alternative is greater or equal than the adaptive decision threshold (Eq. 5), then it means that this alternative should currently be tried.

Let *Deflection Allowed* ($DA$) denotes a boolean variable to select either to deflect or to retransmit the burst.

(6)

$$DA(defl) = \begin{cases} True & \text{if } SP(defl) \geq SP_{th}(BLR_{Topo}, U_{Topo}) \\ False & \text{otherwise} \end{cases}$$

$DA$ function is used to determine if the current burst should be deflected by considering current network conditions.

Obviously, those formulas are pre-calculated periodically and a typical routing table is periodically updated so that the forwarding process is not penalised.

Several techniques could be used to find good weights (in Eq. 5 and Eq. 2) used in AHDR. Probabilistic graphical models have been extensively studied in machine learning [7] and could be used to find good weights. Neural Network is also a good learning model to fine tune an output metric by considering inputs metrics.

*C. AHDR forwarding process*

When a BHP arrives at a core node, the next hop has to be selected from a routing table in the electrical domain in order to reserve bandwidth for the data burst. Here is the approach used for the forwarding process:

- When a control packet is received, the current node is compared to the destination node. If the BHP arrives at the destination, then an ACK is sent to the source.

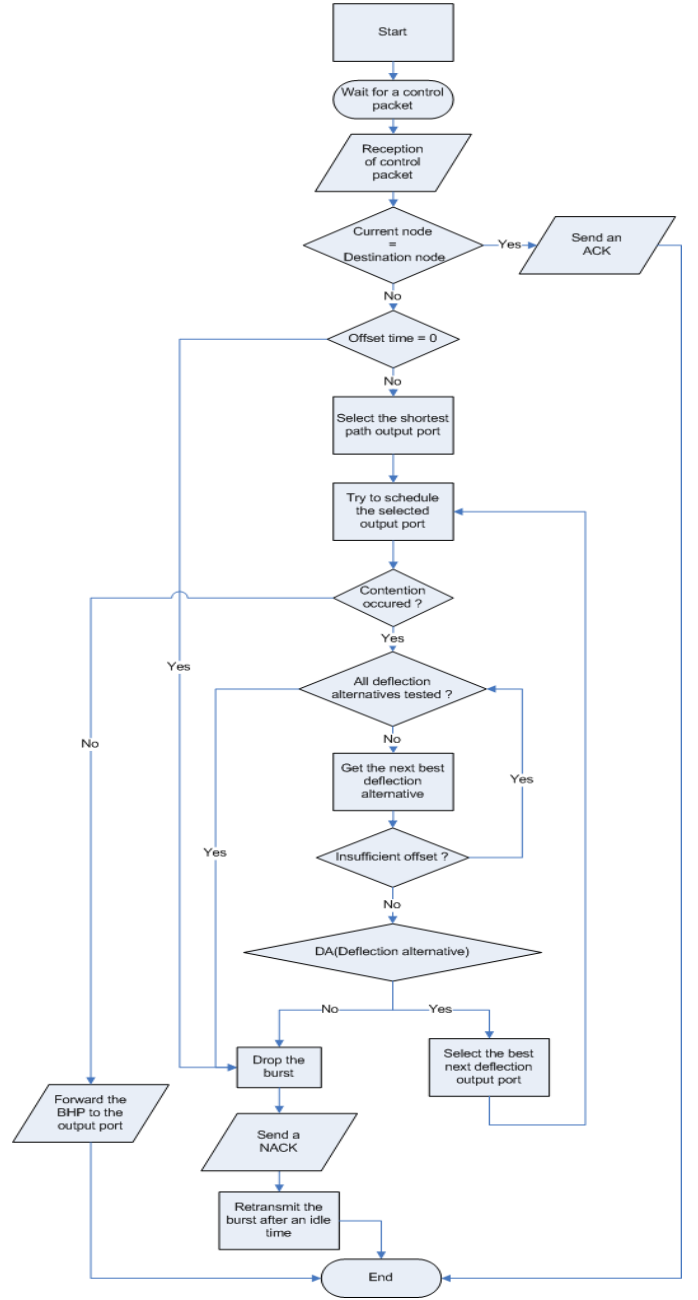

Fig. 6. AHDR forwarding process

- Then, the offset time is checked in order to verify if it is still sufficient. If it is not sufficient, a NACK is sent to the source and the burst is retransmitted after an idle time.
- The shortest path output port is selected. In case of resource contention, it is solved by either deflecting or by retransmitting the burst. The burst is not retransmitted and dropped after a certain number of retransmissions (noted $N_{ret}$).
- AHDR successively extracts best deflection alternatives in order to minimize the BLR and the number of retransmissions. The best output port is found by extracting the next hop in the route $R$ where $SP(R)$ (Eq. 3) is the highest.
- The success probability of the deflection route is then compared to the adaptive decision threshold.

- If the success probability of the current alternative is smaller than the adaptive decision threshold, then a NACK is sent to the source and the burst is retransmitted after an idle time. Otherwise, the current output port alternative is scheduled.

TABLE I
AHDR ROUTING TABLE

| Destination | Next hop | Cost |
|---|---|---|
| $Dest_1$ | $Hop_{1,Dest_1}$ | $1 - SP(R(Hop_1, Dest_1))$ |
| ⋮ | ⋮ | ⋮ |
|  | $Hop_{M,Dest_1}$ | $1 - SP(R(Hop_M, Dest_1))$ |
| ⋮ | ⋮ | ⋮ |
| $Dest_N$ | $Hop_{1,Dest_N}$ | $1 - SP(R(Hop_1, Dest_N))$ |
| ⋮ | ⋮ | ⋮ |
|  | $Hop_{M,Dest_N}$ | $1 - SP(R(Hop_M, Dest_N))$ |

However, to not penalize the forwarding process physically, the routing table (TABLE I) is updated periodically by using Eq. 3 ($SP(Route)$). The cost is expressed by:

$$Cost(Next\ hop, Dest) = 1 - SP(R(Next\ hop, Dest)) \quad (7)$$

where $R(Next\ hop, Dest)$ is the route of a given next hop to the destination so that next hops having a high success probability will result in a low cost. We note that next hops are sorted as follows:

$$\forall_{i=1}^{N} \forall_{j=1}^{M-1} Cost(Hop_j, Dest_i) \leq Cost(Hop_{j+1}, Dest_i)) \quad (8)$$

The decision threshold is also updated periodically by using Eq. 5.

*D. Adaptive offset time*

In OBS networks, the data burst follows the control packet after a predetermined offset time calculated at the ingress node. The offset time has to be large enough so that bursts arrive at each switch after the control packet. The minimum offset time $t_{offset}$ must consider the BHP processing time $t_p$ at each hop, the node switching and the configuration time $t_{conf}$. However, the minimum offset time is expressed by:

$$t_{offset} = t_{conf} + N_{hops} * t_p \quad (9)$$

where $N_{hops}$ is the number of hops. Eq. 9 expresses the fact that the main key to find the best offset time is to predict the number of hops because $t_{conf}$ and $t_p$ are fixed values. However, if deflection occurs, a longer route could be used which increases $N_{hops}$.

The number of hops ($N_{hops}$) to be used in the offset time equation (Eq. 9) is given by:

$$\quad (10)$$

$$N_{Hops} = \begin{cases} |Best\ defl| & \text{if } DA(Best\ defl) \text{ is true} \\ |Shortest\ path| & \text{otherwise} \end{cases}$$

where $|Best\ defl|$ means the path length of the best deflection alternative from the ingress node. If $DA(Best\ defl)$ (from Eq. 6) is true, then the best deflection route is used for the number of hops and otherwise the shortest path is used. Eq. 9 is then used to calculate the offset time with an adapted number of hops.

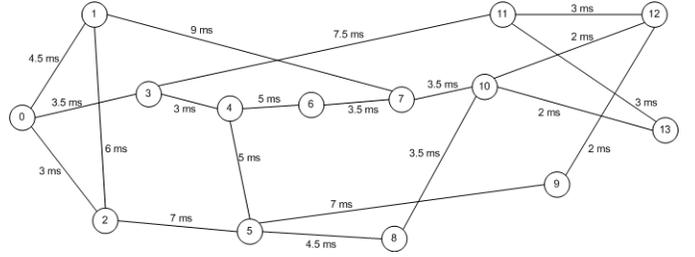

Fig. 7. NSFNET topology

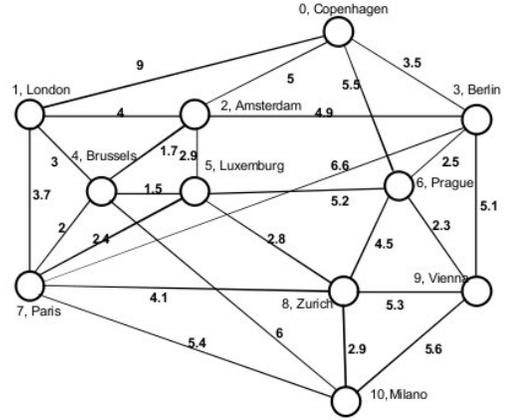

Fig. 8. COST239 topology

III. SIMULATION RESULTS

Simulations are performed with NSFNET (Fig. 7) and COST239 (Fig. 8) topologies by using Network Simulator 2 (ns-2) with an extra module for OBS. Because in LHDR there is only one possible deflection alternative and AHDR try several deflection alternatives, LHDR is modified (MLHDR) in order to do proper comparisons. The only difference between LHDR and MLHDR is that when contention occurs, MLHDR apply LHDR for the shortest alternative first, then for the second shortest alternative and so forth as long as there is a possible deflection alternative.

The following simulation configuration is used:
- Each wavelength has 1 Gbit/s of bandwidth capacity.
- Each link has 2 control channels and 4 data channels.
- The mean burst size equals 400 KB in NSFNET topology and 4 MB in COST239 topology.
- Packet and burst generations follow a Poisson distribution for the input packet rate and for the burst size.
- Traffic generators are distributed randomly over the network. Each traffic generator sends bursts to any node (except himself).
- Bursts are indefinitely lost after a certain number of retransmissions $N_{ret}$ (truncated retransmission). $N_{ret}$ is fixed to 1 in order to not increase significantly the end-to-end delay. Finding the best $N_{ret}$ is out of the scope of this paper.
- Dropped bursts are retransmitted after $Rand(0, 0.05)$ seconds so that retransmitted bursts are highly penalized in terms of end-to-end delay. $Rand$ returns a random value between a minimum and a maximum value.
- We define the traffic load to be the ratio of the total input source nodes throughput over the capacity of the whole network [8].

Weights applied to $SP_{th}$ input metrics were varied (Fig. 9) with several loads. Best results in terms of BLR are obtained by using

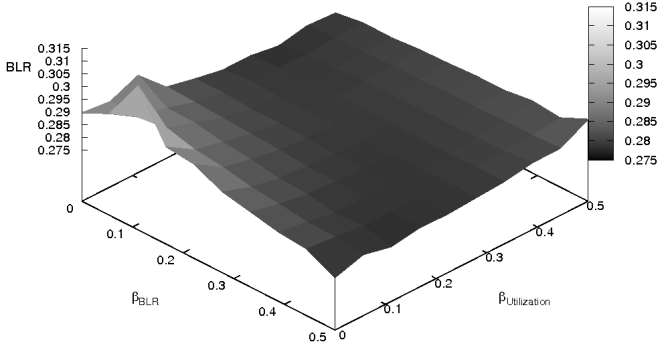

Fig. 9. Weights (used by $SP_{th}$) variations over NSFNET, load = 0.48

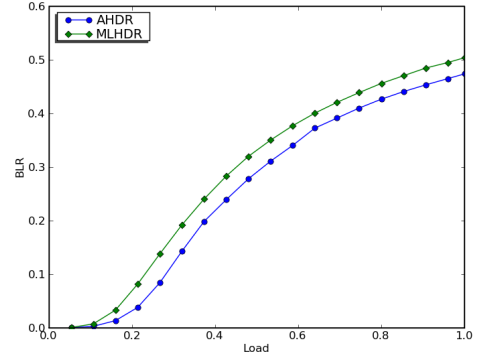

Fig. 10. BLR in NSFNET

weights where $0.4 \leq \beta_{BLR} + \beta_{Utilization} \leq 0.8$.

## A. Comparison of MLHDR and AHDR

In this section, we present the obtained simulation results that compare MLHDR and AHDR performance.

*1) NSFNET topology:* Several simulations were done with NSFNET topology. Let $L$ be the number of links in NSFNET, $N$ the number of nodes, NSFNET has a low connectivity of $C = 0.23$ where $C = \frac{L}{N(N-1)/2}$. AHDR gives significant improvements in terms of BLR even at high loads (Fig. 10). NSFNET has a low connectivity which is a bad case for AHDR since AHDR selects low loaded links. Because of that, when only a few alternatives are available, the gain difference is limited. In this case scenario, the gain comes from two main sources:

- The adapted offset time (Fig. 12) since Eq. 10 is used to adapt the offset time depending on network conditions.
- AHDR forwarding process (see Fig. 6).

We could expect that the end-to-end delay is highly increased because the forwarding process of AHDR can select longer routes compared to the shortest path. However, the end-to-end delay is similar compared to MLHDR. The highest end-to-end delay for AHDR gives 5 ms higher than MLHDR (Fig. 11). From a client (TCP for example) point of view, 5 ms is more than acceptable in general since the Internet uses the *Best Effort* paradigm.

The adapted offset time (see Eq. 9) highly influence the number of deflections ($\# \: deflections$) versus the number of retransmissions ($\# \: retransmissions$) during simulations. We define the deflection ratio as follows:

$$Deflection \: ratio = \frac{\# \: deflections}{\# \: deflections + \# \: retransmissions} \quad (11)$$

Fig. 13 shows the deflection ratio variation of AHDR and MLHDR. With AHDR where $0 \leq Load \leq 0.25$, we can clearly observe that AHDR does as much deflections as possible. However, as the load increases, we can see the benefit of using AHDR compared to a static approach: deflections are done and are effective as long as it reduces the BLR. We can also observe that MLHDR does not enough deflections where $0 \leq Load \leq 0.5$ and does too much deflections where $0.5 < Load \leq 1$.

*2) COST239 topology:* Several simulations were also done in COST239 topology in order to compare AHDR and MLHDR with a highly connected topology ($C = 0.47$). AHDR gives significative improvements in terms of BLR (Fig. 14). For loads between 0 and

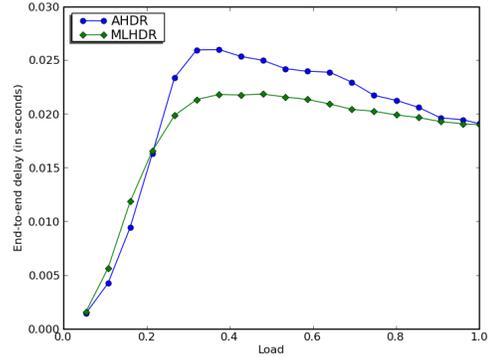

Fig. 11. End-to-end delay in NSFNET

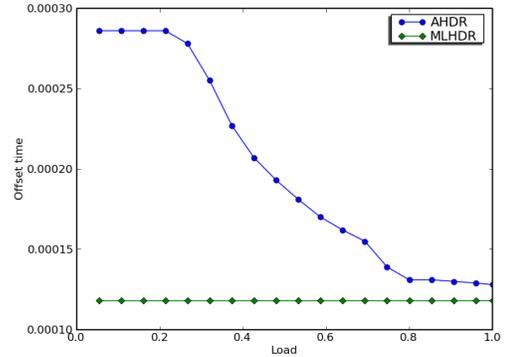

Fig. 12. Offset time variation in NSFNET

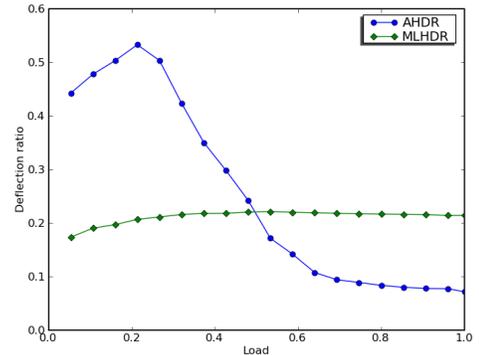

Fig. 13. Deflection ratio in NSFNET

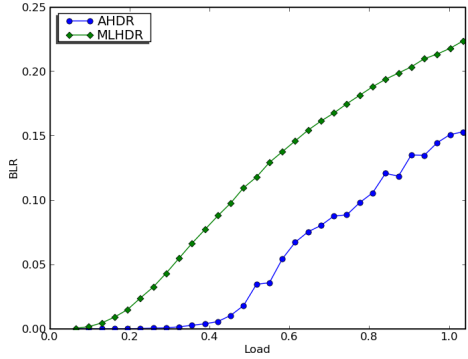

Fig. 14. BLR in COST239

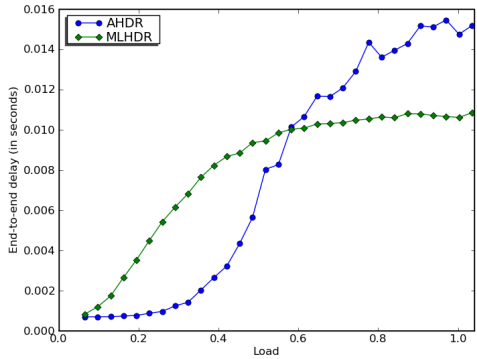

Fig. 15. End-to-end delay in COST239

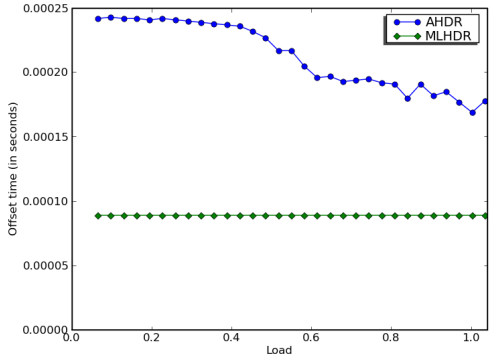

Fig. 16. Offset time variation in COST239

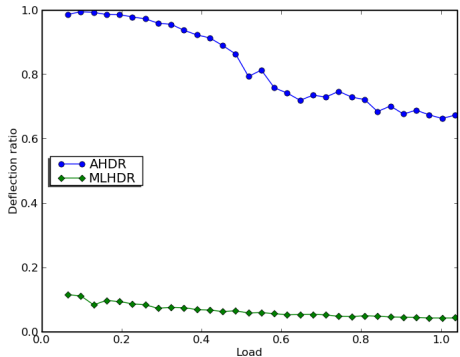

Fig. 17. Deflection ratio in COST239

0.6, there is also a gain in terms of end-to-end delay (Fig. 15). For loads greater than 0.6, the end-to-end delay is maximally 6 ms higher with AHDR in order to reduce the BLR. The deflection ratio should be high with highly connected topologies like COST239 at least when the load is low. Static hybrid deflection and retransmission mechanisms such as MLHDR under-utilize deflection routing (Fig. 17) where the deflection ratio is approximately always 0.08. AHDR always deflect bursts when the load is low. As the load increases, the offset time is reduced (Fig. 16) in order to reduce the deflection ratio in an adaptive manner.

## IV. CONCLUSION AND FUTURE WORK

This paper presents a novel algorithm called *Adaptive Hybrid Deflection and Retransmission* (AHDR) that combines deflection and retransmission routing. In order to make effective decisions between those two contention resolution strategies, AHDR uses an adaptive decision threshold. This decision threshold is adapted using network metrics such as BLR and link utilization since there exists a correlation between the BLR and the best decision threshold to use in order to reduce the BLR. Low connected topologies such as NSFNET needs an adaptive mechanism to balance deflection routing and retransmission in order to reduce the BLR with a small cost in terms of end-to-end delay. Also, static approaches such as MLHDR in NSFNET do not deflect bursts enough when the load is low and does too much deflections as the load increases. Highly connected topologies such as COST239 offers the ability to have a high ratio of deflections over retransmissions in order to reduce the BLR with a small cost in terms of end-to-end delay when the load is high.

The future work of this research is to combine several contention resolution strategies in a dynamic way because we believe that the feasibility of OBS requires effective and adaptive algorithms to overcome the burst loss issue. We are presently working on a new approach which deploys a probabilistic graphical model used in artificial intelligent in order to make efficient and dynamic decisions among several contention resolution strategies.

## ACKNOWLEDGMENT

The authors would like to thank Son-Hong Ngo for answering several of our questions about his paper [5].